\documentclass[journal=jacsat,manuscript=article]{achemso}

\usepackage[version=3]{mhchem} 
\usepackage{prettyref}

\author{Tatsuya Sagawa}
\affiliation{Undergraduate School of Pharmaceutical Sciences, Kyoto University, Japan}
\author{Ryosuke Kojima}
\affiliation{Graduate School of Medicine, Kyoto University, Japan}
\email{kojima.ryosuke.8e@kyoto-u.ac.jp}

\title[ReactionT5]
  {ReactionT5: a large-scale pre-trained  model towards application of limited reaction data}

\abbreviations{DL}

\begin{document}

\begin{abstract}
Transformer-based deep neural networks have revolutionized the field of molecular-related prediction tasks by treating molecules as symbolic sequences.
These models have been successfully applied in various organic chemical applications by pretraining them with extensive compound libraries and subsequently fine-tuning them with smaller in-house datasets for specific tasks. However, many conventional methods primarily focus on single molecules, with limited exploration of pretraining for reactions involving multiple molecules.

In this paper, we propose ReactionT5, a novel model that leverages pretraining on the Open Reaction Database (ORD), a publicly available large-scale resource. We further fine-tune this model for yield prediction and product prediction tasks, demonstrating its impressive performance even with limited fine-tuning data compared to traditional models.
The pre-trained ReactionT5 model is publicly accessible on the Hugging Face platform.
\end{abstract}

\newcommand{\mvec}[1]{\mathbf{#1}}
\newcommand{\mset}[1]{\mathbf{#1}}
\newcommand{\argmax}{\mathop{\rm argmax}\limits}
\newcommand{\argmin}{\mathop{\rm argmin}\limits}
\newcommand{\commentsize}{\footnotesize}
\newcommand{\sitepath}[1]{{\sf #1}}
\newcommand{\action}[1]{{\tt #1}}
\newcommand{\nonterminal}[1]{{\sf #1}}
\newcommand{\method}[1]{{\sf #1}}
\newcommand{\defeq}{\ensuremath{\stackrel{\mathrm{def}}{=}}}

\newcommand{\glabel}[1]{{\sf #1}}

\newrefformat{fig}{Figure~\ref{#1}}
\newrefformat{sec}{Section\ref{#1}}
\newrefformat{tbl}{Table~\ref{#1}}
\newrefformat{eq}{Eq.(\ref{#1})}
\newrefformat{alg}{Algorithm \ref{#1} }

\section{Introduction}

Deep learning models related to organic chemical reactions have gained increasing attention in recent years as significant challenges \cite{coley2017prediction}.
Traditional approaches based on the experience and knowledge of expert chemists often struggle with the complicated and diverse nature of reactions.
Contrastingly, deep learning approaches are expected to support chemists with data-driven analysis on large reaction datasets.

In recent years, deep learning has witnessed significant advancements through the construction of pre-trained models using massive datasets, followed by fine-tuning with limited in-house data, leading to remarkable performance improvements.
This approach has proven successful in various domains, including computer vision, speech recognition, and natural language processing (NLP) \cite{zhou2023comprehensive}.
Particularly, the discovery of ``scaling law'' in deep learning, revealing the potential of training large-scale transformer-based models, has garnered significant attention \cite{scaling2021transfer}.

Against this background, in organic chemistry, large-scale models pretrained by compound libraries have been greatly focused \cite{xia2023systematic}.
Many such methods deal with a molecule as a symbolic sequence like natural language text.
For example, SMILES-BERT has been reported to produce high performance by representing compounds in a Simplified Molecular-Input Line-Entry System (SMILES) format and performing unsupervised pretraining \cite{wang2019smiles}.
Such methods are also developed for generative models of molecules, e.g., MolGPT for molecule generation using a generative pre-trained transformer (GPT), which is known as a language model \cite{bagal2021molgpt} and AlphaDrug for generating target molecules in combination with reinforcement learning \cite{qian2022alphadrug}.
Towards a wide range of applications involving both natural language text and molecules, question-answering tasks on chemistry and molecular captioning have also been tackled using transformer-based models \cite{edwards-etal-2022-translation,christofidellis2023unifying}.

While transformer-based models that use a single molecule as input and/or output are being vigorously developed as described above, the insight of pre-trained models for multiple molecules, including chemical reactions, is still limited.
As one of the advanced efforts, T5Chem supports multitasking reactions such as yield prediction and product prediction \cite{lu2022unified}.
However, T5Chem focuses on building a single model for multitasking and does not address applying pre-trained models to the other smaller datasets.
Our goal is to investigate the effectiveness of a pre-trained model using the large-scale reaction dataset for fine-tuning with in-house data.
We believe that this framework is important to remove discrepancies between the training set and our in-house data over a large reaction space.

We propose a new transformer-based pre-trained model, named ReactionT5, that works well by fine-tuning with a small dataset of organic reactions.
Our model is based on the text-to-text transfer transformer (T5), which accepts text input and text output and is known as text-to-text tasks such as translation and question-answering tasks \cite{raffel2020exploring}.
To pretrain large-scale models, we adopt a two-stage pretraining procedure: compound and reaction pretraining.
In the compound pretraining stage, the initialized T5 model is trained using a single molecule as input and output, and the trained model is named CompoundT5.
In the reaction pretraining stage, CompoundT5 is additionally pretrained using a reaction database.
Once this model is pretrained through these two steps, we can carry out fine-tuning this model on a target dataset, whose size is often relatively small.
In this paper, we demonstrate the effectiveness of our approach by applying this model to yield and product prediction tasks.

\begin{figure*}[th]
\centering
\includegraphics[width=0.9\textwidth]{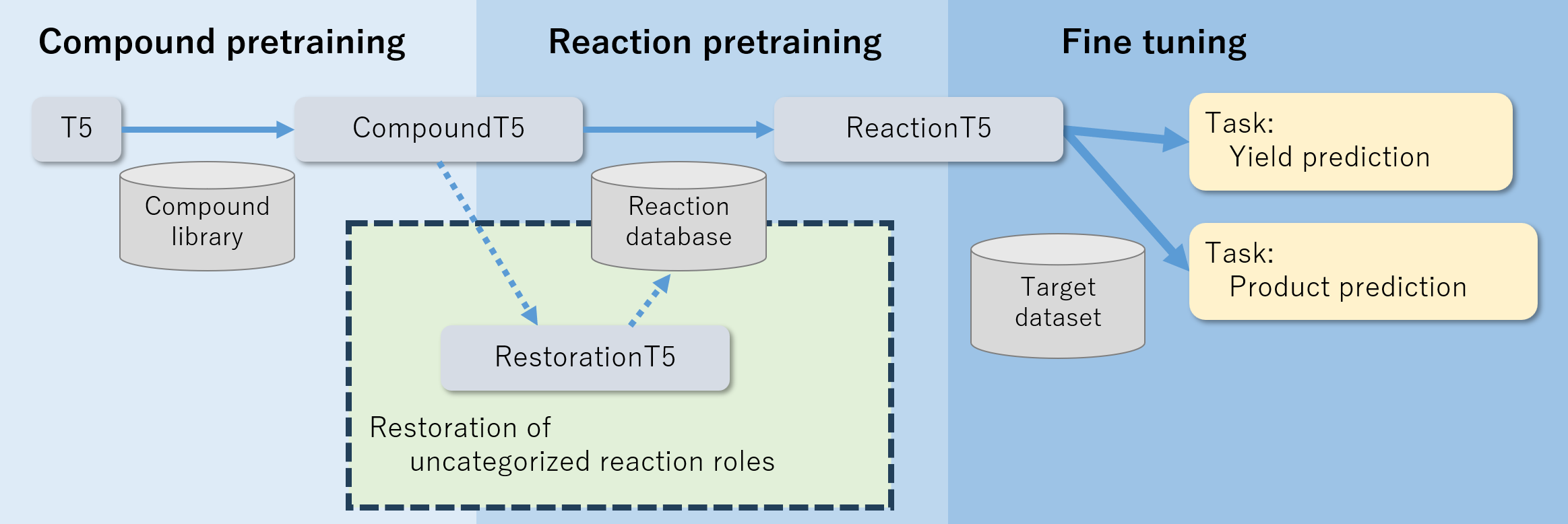}
\caption{
A workflow of our method. We start with the base T5 model, from which we derive CompoundT5 through pretraining on compound pretraining.
Next, we introduce RestorationT5 developed from the CompoundT5 model to restore uncategorized data in the reaction database.
After that, ReactionT5 is constructed using a restored reaction database from CompoundT5.
Finally, fine-tuning ReactionT5 is performed by using relatively little data like in-house data.
}
\label{fig:overview}
\end{figure*}

For the reaction pretraining, we utilize the open reaction database (ORD), an open-access platform for organic reaction data \cite{ord2021}.
Because ORD aggregates a wide range of reaction data containing benchtop reactions, automated high-throughput experiments, and flow chemistry, this dataset is suitable for pretraining our models from the viewpoint of the dataset size and reaction variations.
We aim to construct a pretraining model for organic reactions by pretraining this large-scale reaction dataset extracted from ORD.
This pre-trained model is available via Hugging Face \footnote{ReactionT5 for product prediction : \\
\hspace*{2em} \url{https://huggingface.co/sagawa/ReactionT5-product-prediction} \\ \hspace*{1.5em} ReactionT5 for yield prediction: \\ 
\hspace*{2em} \url{https://huggingface.co/sagawa/ReactionT5-yield-prediction}}.

\begin{figure*}[th]
\centering
\includegraphics[width=0.9\textwidth]{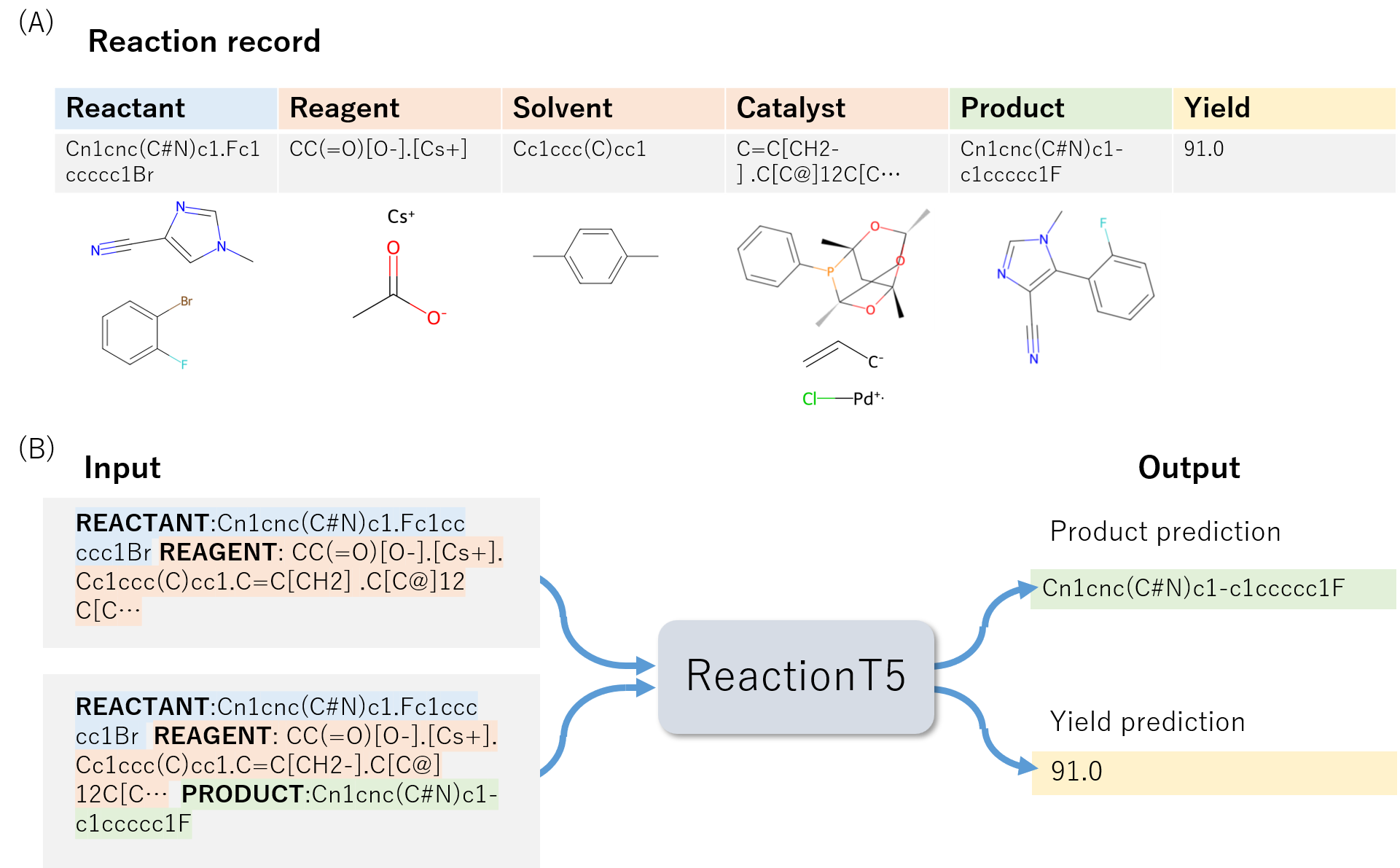}
\caption{(A)  A reaction record in the reaction database is represented as a reactant, reagent, solvent catalyst, product, and yield. All the compounds in this record are represented in SMILES format. (B) ReactionT5 uses text format for inputs, where SMILES and special tokens representing reactant, reagent, and product, to output the product or yield of the reaction.}
\label{fig:t5}
\end{figure*}

\section{Method}

For the base model, this study employs T5, which consists of an encoder converting an input text into the hidden vectors and a decoder to construct an output text from those hidden vectors.
We formulate the tasks related to reactions as a form of text-to-text by treating a chemical reaction as a text.
More concretely, we tackle product and yield prediction by formulating those tasks as text-to-text conversion.

\prettyref{fig:overview} shows an overview of our method.
Our method involves two-stage pretraining processes, which comprise compound pretraining and reaction pretraining.
In the stage of compound pretraining, we train both the tokenizer and T5 using a compound library.
The purpose of this stage is to train the T5 model with a single molecular structure contained in the compound library.
We call a T5 model pretrained in this stage CompoundT5.
During the subsequent reaction pretraining stage, CompoundT5 continues to be pretrained by using chemical reaction data containing multiple compounds such as reactants, products, and catalysts.
We call the dataset used in this stage reaction pretraining dataset, and the pre-trained model is called ReactionT5.
Once ReactionT5 has been pre-trained through these two stages, it can be expected to be fine-tuned to better adapt to the domain of the target dataset.
We assume that the data size of this target data is limited as is often the case in real applications.
Details of each pretraining stage are described in the following two sections.

\subsection{Compound pretraining}

In the first stage, we construct a CompoundT5, pretrained from an initialized T5 using a set of single compounds represented in the SMILES format.
To realize this pretraining, we adopt span-masked language modeling, a self-supervised learning task where a sequence of multiple tokens is replaced with a mask token, in contrast to traditional masked language modeling, which replaces only a single token \cite{raffel2020exploring}.
The objective of this task is to predict the original token from the masked tokens.

Our experimental setup to train CompoundT5 basically follows the original T5 \cite{raffel2020exploring}. 
Specifically, we randomly select and replace 15\% of the tokens in the input text with mask tokens, with the average length of replaced tokens being 3.
We use a SentencePiece unigram tokenizer, trained using compound library \cite{kudo2018sentencepiece}. This tokenizer can represent compounds with fewer tokens compared to character-level or atom tokenizers, facilitating faster training and inference, as well as enabling the prediction of larger compounds.

We chose ZINC as the compound library for this pre-training.
As a preprocessing at this stage, all SMILES in this dataset are canonicalized by using RDKit.
More details are described in the Datasets section.



\subsection{Reaction pretraining}

In the second stage, we conduct reaction pretraining to address two tasks: compound and yield predictions.
These tasks are designed as a text-to-text representation, i.e., inputs and outputs (products or yields) are represented as texts based on the SMILES format (\prettyref{fig:t5} top).
In this stage, both two tasks use a reaction pretraining database where one record representing the reaction consists of six roles: reactant, reagent, solvent, catalyst, product, and yield.
The field of yield is in numerical form, otherwise, zero or more molecules are expressed in the SMILES format.
To represent these six roles, six special tokens are added to the tokenizer used in the CompoundT5.



\subsection{Product prediction}

In the product prediction task, the T5 model aims to output text containing the SMILES of the products from the input text consisting of the reactants, catalysts, reagents, and solvents.
In this setting, catalysts, reagents, and solvents are categorized into reagents because it is difficult to distinguish between these three due to different annotation rules for each dataset.
Using special tokens, ‘REACTANT:’ and ‘REAGENT:’, added before the SMILES of reactant and reagent, respectively, information in the input is represented as a single piece of text.
These special tokens are added to the tokenizer before training.

One of the important parts of T5 that determines the quality of the output text is the decoder whose last layer computes a vector consisting of probability distribution on the vocabulary.
The model repeatedly predicts output tokens using the input into the decoder and a hidden vector from the encoder.
The decoder of the model has the same vocabulary as the encoder, and its output is a probability distribution over all tokens.
Sequence-based models predict a probability distribution over all tokens during text generation and iteratively select the most probable token as the next token and end prediction when an end-of-sequence (EOS) token appears.
The probability of each text is calculated by multiplying the chosen tokens’ probability.



In this study, we set minimum and maximum lengths for product prediction.
By setting these lengths, we ensure the model continues prediction until token lengths reach the minimum length and halt prediction if token lengths reach the maximum length.
We observed that this setting significantly influences the model’s performance.
Therefore, we conducted a search for the best practice in setting these minimum and maximum lengths (Appendix 2).
Additionally, product prediction using the T5 models utilizes beam search to calculate the most likely token sequence, where the size of the beam affects performance.
In this study, we set the beam size to 10 (Appendix 3).

\subsection{Yield prediction}

For the yield prediction task, every record representing a reaction in the reaction dataset is preprocessed to create input and output sequences as shown in \prettyref{fig:t5} (bottom).
When the dataset contains records with yields greater than 100, we clip them to 100 and normalize all yields to fall within a range from 0 to 1.
An input token sequence representing a single reaction consists of the reactants, catalysts, reagents, and products.
Similar to the product prediction task, the categories of catalysts, reagents, and solvents are concatenated using a period and treated as a single reagent.
We add special tokens, ‘REACTANT:’, ‘REAGENT:’, and ‘PRODUCT:’, before the reactant, reagent, and product SMILES sequence, respectively.
These are then concatenated into a single reaction. These special tokens are added to the tokenizer before training.
As a loss function of yield prediction, we utilize mean squared error loss for numerical output.


\section{Datasets}

\begin{figure}[t]
    \begin{tabular}{cc}
        \begin{minipage}{.5\textwidth}
            \centering
        \includegraphics[width=0.9\textwidth]{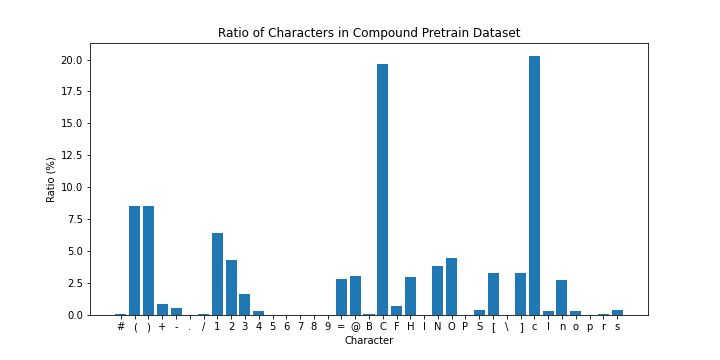}
        \end{minipage}
        \begin{minipage}{.5\textwidth}
            \centering   \includegraphics[width=0.9\textwidth]{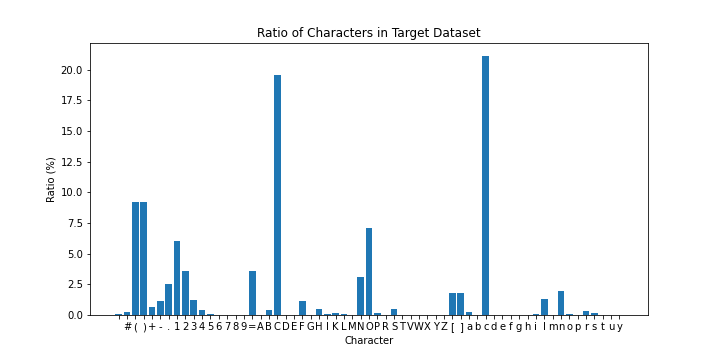}
        \end{minipage}
    \end{tabular}
\caption{Comparison of counts of characters appearing in Compound pretrain dataset and Target dataset. Target dataset contains a greater variety of characters compared to Compound pretrain dataset. }
\label{fig:ratio}
\end{figure}

\subsection{Compound library}

We use 23 million compounds from the ZINC database \cite{irwin2020zinc20}, where each molecule is represented in the SMILES format, as the compound library.
\prettyref{fig:ratio} (left) shows the number of characters appearing in this dataset.
The experimental settings for this pre-training stage follow those for T5Chem \cite{lu2022unified}.

\subsection{Reaction pretraining database: ORD}

\begin{figure}[t]
\centering
\includegraphics[width=0.45\textwidth]{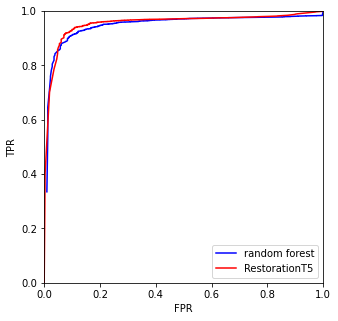}
\caption{ROC curve of restoration related to uncategorized roles of compounds in the ORD}
\label{fig:uncategorized}
\end{figure}

\begin{table}[t]
  \caption{Performance of restoration related to uncategorized roles of compounds in the ORD}
  \label{tbl:uncategorized}
  \centering
  \begin{tabular}{l|rrr}
    \hline
Model	& precision	& recall	& F1 score \\ \hline
RestorationT5	& 0.0878	& 0.7212	& 0.1564 \\
random forest	& 0.0663	& 0.3942	& 0.1136 \\
    \hline
  \end{tabular}
\end{table}

We utilize 1.5 million reactions from the ORD as the reaction pretraining database.
The ORD consists of data from various sources, and some of the data includes atom mapping information.
In real-world applications, atom mapping is often inaccessible when employing the trained model for reaction prediction.
So, using RDKit, we removed information on atom mapping from ORD so that the model would be independent of the given atom mapping.
Additionally, we canonicalized all compounds in the ORD.
Although the original ORD contains additional information on molecules in the reactions, we only use the reaction roles such as reactants, reagents, solvents, catalysts, internal standards, and products. We classified the remaining compounds as uncategorized.


\prettyref{fig:ratio} (right) shows the number of characters appearing in the ORD.
Comparing these left and right figures, we found that the ORD consists of a more diverse set of characters than the ZINC database.
From these facts, a tokenizer trained on the pre-training dataset cannot recognize specific atoms present in the data in the ORD.
Specifically, metal atoms, which influence reaction progression and yields as catalysts, are not recognized, leading to a degradation in the model’s performance.
Therefore, after compound pretraining, we add new tokens to the tokenizer to ensure it covers all tokens in the reaction dataset.

The specific roles of compounds in real-world chemical reactions, such as reactants, reagents, and catalysts, are often unknown.
For example, the ORD contains 479,035 reactions with such 'uncategorized' compounds, which is 31.81\% of all reactions (see Appendix 1).

Our preliminary investigation found that the 1.8 million reactions without such uncategorized roles in the ORD had a high bias related to products because only 447 unique compounds appeared in such reactions.
In contrast, the rest of the reactions with uncategorized compounds contained 439,898 unique product compounds.
Given these facts, simply removing reactions with missing reaction roles from the dataset may lead to bias in the product prediction model.


In light of this, to increase the amount of available training data by utilizing the uncategorized compound in the ORD, we preliminary addressed the task of category restoration.
More concretely, this task is formulated as a binary classification of an uncategorized compound into a reactant and reagent.
To address this task, we developed another classifier based on CompoundT5 for uncategorized compounds, referred to as RestorationT5. The architecture of RestorationT5 is largely similar to the model used for yield prediction, but we modified the last output layer to the sigmoid function for binary classification.
This alteration enables the model to classify uncategorized compounds as reactants (negative labels) or reagents (positive labels).
For evaluation, we compared RestorationT5 with a random forest model that uses Morgan fingerprints as input features.
Given the imbalanced nature of the target data, we oversampled the reagent data 100 times to stabilize the model training. 
\prettyref{fig:uncategorized} illustrates the receiver operating characteristic (ROC) curve obtained by adjusting the threshold value.
We selected 0.97 as the optimal threshold, as it provides the best F1 score, and proceeded to restore uncategorized compounds.
\prettyref{tbl:uncategorized} compares the performance of RestorationT5 and the random forest model, with RestorationT5 showing superior performance.
Therefore, we use a new dataset restored by RestorationT5 as ORD(restored) in later experiments.

\subsection{Target dataset}


We utilize the USPTO dataset \cite{coley2017prediction}  as the target dataset for product prediction.
The USPTO dataset is based on Lowe’s patent data \cite{lowe_2012} and contains 479,000 reaction data.
Following the previous study \cite{lu2022unified}, we split the data into train, validation, and test sets (comprising 409,000, 30,000, and 40,000 data, respectively). 
Since 18\% of reactions in the USPTO dataset were included in the ORD, these reactions were removed during evaluation.

We utilize the palladium-catalyzed Buchwald-Hartwig C-N cross-coupling reactions dataset for yield prediction.
This dataset, created based on Ahneman et al.’s high-throughput experiment \cite{ahneman2018predicting}, contains 3,955 Buchwald-Hartwigh cross-coupling reactions of heteroaryl halides and 4-methylaniline catalyzed palladium with various reaction inhibitors.
We confirmed that the reactions in this dataset were not included in the ORD.

Consistent with previous research, we performed the experiment on the Buchwald-Hartwig C-N cross-coupling reactions dataset 10 times by randomly dividing the training and testing datasets into a 7:3 ratio.
This dataset also contains four out-of-sample datasets to assess the model’s performance (Test 1-4).
We use the $R^2$ metric for the evaluation of yield prediction.
This metric measures the correlation between forecast and target values, so closer to 1 means better performance.

\section{Result}

In this section, we evaluate ReactionT5 pre-trained in ZINC and ORD by applying this model to product prediction and yield prediction tasks.

\subsection{Product prediction}

\begin{table*}[t]
  \caption{Top-k accuracy and invalidity (\%) of CompoundT5, ReactionT5, and conventional models in product prediction. }
  \label{tbl:product_finetune_uspto}
  \centering
  \begin{tabular}{lllrrrrr}
    \hline
	& Train	& Test	& Top1	& Top2	& Top3	& Top5 & invalidity\\ \hline
Seq-to-seq	& USPTO	& USPTO	& 80.3	& 84.7	& 86.2	& 87.5 & - \\
WLDN	& USPTO	& USPTO	& 85.6	& 90.5	& 92.8	& 93.4 & - \\
Mol Transformer	& USPTO	& USPTO	& 88.8	& 92.6	& –	& 94.4 & - \\
T5Chem	& USPTO	& USPTO	& 90.4	& 94.2	& –	& 96.4 & - \\
CompoundT5	& USPTO	& USPTO	& 88.0	& 92.4	& 93.9	& 95.0 & 7.5 \\
ReactionT5(ORD)	& -	& USPTO	& 0.0	& 0.0	& 0.0	& 0.0  & 0.6\\
ReactionT5(ORD)	& USPTO200	& USPTO	& 0.0	& 0.0	& 0.0	& 0.0  & 4.2\\
ReactionT5(restored ORD)	& -	& USPTO	& 0.0	& 0.0	& 0.0	& 0.0  & 1.1\\
ReactionT5(restored ORD)	& USPTO200	& USPTO	& 85.5	& 91.7	& 93.5	& 94.9  & 12.0\\
    \hline
  \end{tabular}
\end{table*}

\begin{table}[t]
  \caption{Top-k accuracy and invalidity (\%) when varying the number of samples from USPTO for fine-tuning}
  \label{tbl:product_finetune_small}
  \centering
  \begin{tabular}{rrrrrr}
    \hline
\#sample	& Top1 	& Top2 	& Top3 	& Top5 	& invalidity  \\ \hline
10	&  9.0	& 12.5	& 15.3	& 19.1	& 12.4 \\
30	& 80.5	& 87.3	& 89.8	& 92.0	& 17.2 \\
50	& 83.7	& 89.9	& 92.2	& 94.0	& 14.8 \\
100	& 85.1	& 91.0	& 92.8	& 94.4	& 14.0 \\
200	& 85.5	& 91.7	& 93.5	& 94.9	& 12.0  \\
    \hline
  \end{tabular}
\end{table}

In product prediction, we utilized the USPTO dataset for evaluation.
We generated the top-1 to top-5 predicted products by using probability scores computed in the decoder of T5 and computed the top-1 to top-5 accuracy.
When the true SMILES is included in the top-k predictions, it is considered correct.
In the SMILES comparison here, only if the strings after the canonical match exactly are counted as a match.
Note that the output SMILES is canonicalized before comparison.
Also, we define predictions as invalid if they could not be recognized by RDKit, and calculated invalidity as the proportion of invalid predictions in the top-5 predictions.

\prettyref{tbl:product_finetune_uspto} shows that the benchmark performance of our ReactionT5 against conventional methods: a sequence-to-sequence model (Seq-to-seq) \cite{schwaller2018found},
a Weisfeiler-Lehman Difference Network (WLDN) based on a graph neural network called Weisfeiler-Lehman Network (WLN) \cite{lei2017deriving,jin2017predicting},
a transformer-based model for reaction prediction (Mol Transformer) \cite{schwaller2019molecular} \footnote{Regarding WLDN, the different performance is reported between the original paper and \cite{lu2022unified}, but here we adopted the original paper with higher performance.},
a T5-based model called T5Chem, \cite{lu2022unified}.
ReactionT5 pre-trained on the ORD with and without restored uncategorized roles exhibited zero accuracy on the ReactionT5(ORD) and ReactionT5(restored ORD) rows of this table.
We attribute this to a discrepancy between the training and test datasets, e.g., the ORD reactions often include byproducts as their products, while the USPTO reactions focus solely on the main product.


To address this issue, we fine-tuned our ReactionT5 using 200 reactions from the USPTO training data, a small fraction of the original 409,000 reactions (USPTO200 in \prettyref{tbl:product_finetune_uspto}).
This process minimally affected the ReactionT5 trained on the ORD but substantially improved its performance when combined with the restoration of uncategorized compounds from the ORD.
The resultant model demonstrated competitive performance against other models trained on the full dataset despite using only 200 reactions from the target USPTO dataset.

We further investigate in more detail the performance improvement effect of fine-tuning ReactionT5 using the small amount from the USPTO training dataset mentioned above.
\prettyref{tbl:product_finetune_small} shows that the relationship between the number of reactions used in the fine-tuning from USPTO and the performance of the fine-tuned ReactionT5.
In this experiment, we used a pre-trained model with the restored ORD dataset and fine-tuned it by varying the number of data reactions from 10 to 200.
This result shows that it is possible to achieve Top1 accuracy of more than 80\% by using fine-tuning with more than 30 reactions from USPTO.
Consequently, this result shows that a model pre-trained with the restored ORD can be adapted by fine-tuning a limited amount of reactions in the target domain dataset.

\subsection{Yield prediction}

\begin{table*}[t]
  \caption{Performance comparison of CompoundT5, ReactionT5, and other models in yield prediction.}
  \label{tbl:yield_result_cn}
  \centering
  \begin{tabular}{lrrrrrr}
    \hline
$R^2$&  Random 7:3& Test 1& Test 2& Test 3& Test 4& Avg. Tests 1–4 \\\hline
 DFT&   0.92&  0.80&  0.77&  0.64&  0.54&  0.69 $\pm$ 0.104 \\
 MFF&   0.927 $\pm$ 0.007&  0.851&  0.713&  0.635&  0.184&  0.596 $\pm$ 0.251 \\
 Yield-BERT&  0.951 $\pm$ 0.005&  0.838&  0.836&  0.738&  0.538&  0.738 $\pm$ 0.122 \\ 
 T5Chem&  0.970 $\pm$ 0.003&  0.811&  0.907&  0.789&  0.627&  0.785 $\pm$ 0.094  \\
 CompoundT5& {\bf 0.971 $\pm$ 0.002}&  0.855&  0.852&  0.712&  0.547&  0.741 $\pm$ 0.126   \\
 ReactionT5& 0.966 $\pm$ 0.003&  0.914&  {\bf 0.940}&  0.819&  0.896&  0.892 $\pm$ 0.045 \\
 ReactionT5  & 0.904 $\pm$ 0.001 &  {\bf 0.919} &  0.927 &  {\bf 0.847} &  {\bf 0.909} &  {\bf 0.900 $\pm$ 0.031} \\
  (zero-shot)  &&&&&&\\
    \hline
  \end{tabular}
\end{table*}

For yield prediction, we use the palladium-catalyzed Buchwald-Hartwig C-N cross-coupling reactions dataset.
This dataset consists of a 7:3 dataset for training and evaluation using a random split, and four test datasets for external evaluation.
In Tests 1-4, datasets were partitioned such that similar reactions were not included in both the train and test sets.
These tests demand models to exhibit effective generalizability to score well.

\prettyref{tbl:yield_result_cn} shows the coefficient of determination ($R^2$) and the root mean squared error (RMSE) of ReactionT5 against models proposed in the previous research.
In this table, we compared our models with four conventional methods: a random forest model with descriptors associated with the reaction for yield prediction (DFT) \cite{ahneman2018predicting}, 
a random forest model that uses multiple fingerprint features (MFF) \cite{sandfort2020structure}, a BERT-based model consisting of an encoder transformer model and a regression layer (Yield-BERT) \cite{Schwaller_2021}, and the T5Chem model, created for multitask reaction prediction.

From this result, CompoundT5 fine-tuned on the C-N cross-coupling reactions in the training dataset, trained with almost the same settings as T5Chem,  shows the best performance in the random splitting dataset.
Its performance was comparable to those reported in previous research, but it was less successful on Tests 1-4.
In contrast, ReactionT5 trained on ORD, showed slightly lower performance on the Random 7:3 but excelled on Tests 1-4.
The difference was most significant in Test 4, which is the most difficult test in comparative methods such as T5Chem.
ReactionT5 consistently made the most accurate predictions across all Tests 1-4, demonstrating that pretraining with ORD significantly enhanced the model’s generalizability.
While ReactionT5’s performance was slightly lower than that of other models when tested against Random 7:3 data, it showed high performance when faced with the more challenging Tests 1-4.

\begin{table}[t]
  \caption{The result of yield prediction with 30\% and 70\% reactions of 30\% the entire C-N cross coupling reactions dataset}
  \label{tbl:yield_result_cn_small}
  \centering
  \begin{tabular}{llllrr}
    \hline
model	& Train	& Validation	& $R^2$	& RMSE	\\ \hline
Random forest	& 30\%	& 70\%	& 0.853	& 10.390	\\
ReactionT5	& 30\%	& 70\%	& {\bf 0.927}	& {\bf 7.330}	\\
ReactionT5 (zero-shot)	& -	& 70\%	& 0.898	& 8.693	\\ \hline
Random forest	& 70\%	& 30\%	& 0.924	& 7.581	\\
ReactionT5	& 70\%	& 30\%	&  {\bf 0.969}	& {\bf 4.861}	\\
    \hline
  \end{tabular}
\end{table}

To evaluate ReactionT5 in cases with less data for fine-tuning, we adjusted the training-to-test ratio to 30:70 to simulate a scenario where limited data is available for training (\prettyref{tbl:yield_result_cn_small}).
In situations with limited training data, ReactionT5 outperformed the random forest model with the Morgan fingerprint.
Moreover, the performance of ReactionT5 was comparable to that of a random forest trained with a 70\% dataset.
Surprisingly, we also found that zero-shot ReactionT5, which is not fine-tuned for C-N coupling, shows higher performance than the random forest trained with 30\% of the dataset.

\section{Conclusions}

In this study, we constructed a transformer-based pre-trained model, ReactionT5, using the open reaction database (ORD), a publicly available large-scale reaction database. 
We conducted the two-stage pretraining for this model with the compound library and reaction databases, which resulted in enhanced performance in the yield and product prediction tasks with fine-tuning.
Also, this model demonstrated excellent performance in predicting product and yield, even in zero-shot and/or fine-tuning scenarios.
Through our experiments, we found that restoration of uncategorized compounds in the pretraining phase enhanced the performance of the model in product prediction with minimal data fine-tuning.
We believe that this public pre-trained ReactionT5 can be used for many real-world applications by fine-tuning this model with in-house data.


\section*{Conflicts of interest}
There are no conflicts to declare.

\section*{Acknowledgements}
This work was supported by JSPS KAKENHI Grant 21H05207, 21H05221 (Digi-TOS) and JST Moonshot R\&D Grant Number JPMJMS2021.




\bibliography{rsc} 
\bibliographystyle{rsc} 

\section*{Appendix 1: Dataset}

\prettyref{tbl:stat} shows the basic statistics of the datasets used in our experiment.
Because ZINC is used as a compound library, the number of samples means that of compounds. 
On the other hand, a sample in the the other datasets represents a reaction.
In the reaction databases, the averages molecular weights are calculated using all molecules in all reactions, and the averages of the input token lengths represents the token length per one reaction.

\begin{table*}[th]
  \caption{Statistics in the dataset}
  \label{tbl:stat}
  \centering
  \begin{tabular}{lrrr}
    \hline
	& \#sample	& average molecular weight	& average input token length \\ \hline
ZINC	& 22,992,522	& 337.05	& 39.55 \\
USPTO	& 479,035	& 631.04	& 66.07 \\
C-N	& 3,955	& 1881.91	& 206.32 \\
ORD with yield & 680,043	& 918.83	& 109.15 \\
ORD & 1,074,313	& 676.78	& 72.45 \\
ORD(restored) & 1,505,916	& 666.22	& 70.87 \\
    \hline
  \end{tabular}
\end{table*}

\section*{Appendix 2: Ablation study of the product prediction}

\begin{table*}[th]
  \caption{The scores varied significantly depending on the minimum and maximum lengths set for the output. When we used the minimum and maximum lengths of products in the USPTO train data to set the length constraints for product generation, the model achieved optimal accuracy and invalidity. In this tanble, $L$ is the length of reactant. }
  \label{tbl:length}
  \centering
  \begin{tabular}{llrrrrr}
    \hline
Minimum length	& Maximum length	& Top-1	& Top-2 	& Top-3 	& Top-5 	& invalidity \\ \hline
–	& –	& 12.0	& 12.9	& 13.3	& 13.6	& 82.20 \\
$L$ - 5	& $L$ + 5	& 74.3	& 77.6	& 78.8	& 79.5	& 15.24 \\
$L$ - 10 & $L$ + 10	& 74.3	& 77.6	& 78.8	& 79.5	& 15.24 \\
$L$ - 20 & $L$ + 20	& 74.3	& 77.6	& 78.8	& 79.5	& 15.24 \\
25th percentile & 75th percentile& 45.8	& 48.3	& 49.2	& 50.0	& 33.26 \\
Minimum & Maximum & {\bf 88.0}	& {\bf 92.4}	& {\bf 93.9}
	& {\bf 95.0}	& {\bf 7.46} \\
    \hline
  \end{tabular}
\end{table*}

As described in the methods section, setting appropriate minimum and maximum lengths for text generation is critical, and these parameters significantly influenced the results of product prediction.
We trained CompoundT5 using the USPTO dataset and evaluated its performance under various minimum and maximum length conditions (\prettyref{tbl:length}). Initially, without setting length limits—meaning the model continue predicting tokens until an EOS token was generated—the accuracy was extremely low and the rate of invalid prediction was high.
Across multiple conditions, the optimal accuracy and invalidity were obtained when the minimum and maximum lengths of the products in the USPTO train data were used for the minimum and maximum lengths of product generation.

The beam size in the beam search for the T5 decoder was a robust parameter in our setting.
\prettyref{tbl:beam} shows the top-k accuracies due to changes in the beam size. 
To perform calculations even with a small size of beams, we calculate and compare the Top1 invalidities.

\begin{table*}[th]
  \caption{Top-k accuracy and top-1 invalidity (\%) in product prediction by changing the beam size}
  \label{tbl:beam}
  \centering
  \begin{tabular}{rrrrrr}
    \hline
$\#$ beams	& Top1 	& Top2 	& Top3	& Top5	&  invalidity \\ \hline
1	& 54.4	& –	& –	& –	& 28.5 \\
2	& 54.3	& 65.9	& –	& –	& 19.5 \\
3	& 54.0	& 65.8	& 70.9	& –	& 17.5 \\
5	& 53.8	& 64.9	& 70.2	& 76.7	& 18.0 \\
10	& 53.6	& 64.2	& 68.9	& 74.8	& 17.5 \\
    \hline
  \end{tabular}
\end{table*}

\end{document}